\documentstyle[fleqn,epsfig,11pt,titlepage]{article}

\newcommand{\be}{\begin{equation}}
\newcommand{\ee}{\end{equation}}
\newcommand{\bea}{\begin{eqnarray}}
\newcommand{\eea}{\end{eqnarray}}

\newcommand{\AmS}{{\protect\the\textfont2

  A\kern-.1667em\lower.5ex\hbox{M}\kern-.125emS}} 

\begin{document}

\begin{titlepage}

\begin{center}

 {\Large \bf  Relations for Direct CP asymmetries \\in
  $B\to \rm PP$ and $B\to \rm PV$ decays\footnote{  
Talk given  at the  QCD@Work 2005, Conversano, Bari, Italy, 16-20 June 2005}}
\vspace{0.5cm}

{\large \bf  T. N. Pham\\}

\vspace{0.5cm}

{\it Centre de Physique Th\'eorique, \\
Centre National de la Recherche Scientifique, UMR 7644, \\  
Ecole Polytechnique, 91128 Palaiseau Cedex, France}

\end{center}
\centerline{\large \bf ABSTRACT}

The presence of additional strong phase  from power corrections
and other chirally enhanced terms makes it more difficult to 
predict direct CP asymmetries in  two-body charmless $B$ decays. 
In this talk, I would like to report on a recent work 
on QCD Factorisation and Power Corrections in Charmless B Decays. 
Using the measured branching ratios for $B\to \rm PV$, it is shown that  
power corrections in charmless B decays are probably large, at least 
for penguin dominated $\rm PV$ channels. Since the tree-penguin 
interference responsible for direct CP asymmetries in two-body 
charmless $B$ decays are related by CKM factors and $SU(3)$ symmetry,
we find that, if power corrections other than the 
chirally enhanced power corrections and annihilation topology 
were negligible, QCD Factorisation  would
predict the direct CP asymmetry of $B \to \pi^+ \pi^-$ to be 
about 3 times larger than that of $B \to \pi^\pm K^\mp$, with opposite sign,  
in agreement with the latest measurement from Belle. Similar relations 
are also given for direct CP asymmetries in $B\to\rm PV$ .  

\end{titlepage}

\section{Introduction}
The large  direct CP asymmetries  observed at
BaBar and Belle in $B \to K\pi$ and $B \to \pi\pi$ 
decays~\cite{BaBar,Belle}  indicate a large strong phase in these 
decay amplitudes. Since in general 
QCD Factorisation(QCDF) predicts  a much smaller strong phase and 
 a small CP asymmetry, one would then need  
 important power correction terms  or other power-suppressed
 non-factorisable term (e.g FSI effects etc.) to generate  a 
large strong phase and 
a large CP asymmetry in these decays. In fact the charmed meson
inelastic rescattering FSI effects\cite{Isola,Isola1} or 
charming penguin\cite{Ciuchini} are able to produce a large absorptive 
part and therefore a large strong phase for $B\to PP$ and $B\to PV$ 
amplitudes. Though the presence of these power-suppressed terms 
makes it difficult to predict the amount of CP asymmetry, 
because of $SU(3)$ symmetry and the CKM factor, one can derive however
relations between CP asymmetries in $B\to PP$ and $B\to PV$. In the 
following I shall first present an analysis showing possible evidence
for  power-suppressed terms  in charmless $B$ decays.

In  QCD Factorization (QCDF)\cite{QCDF}, the $O(1/m_{b})$
 power corrections in  penguin matrix
elements and other chirally enhanced corrections could make important
contributions to the penguin-dominated charmless $B$ decays as in 
 $B \to \pi K$ decays.
Other power corrections terms such as annihilation contributions 
may also be present in $\rm PP$ and $\rm PV$ decays as first noticed 
in the perturbative QCD method for charmless $B$  decays\cite{pQCD}
and indicated by
 recent  analysis of charmless two-body non-leptonic
$B$ decays\cite{Du,Aleksan,Cottingham}. 
In a recent work\cite{Zhu}, we have shown that in QCDF, it is possible 
to consider certain
ratios of the $B \to PV$ amplitudes which depend only on the Wilson
coefficients and the known hadronic parameters. The discrepancy
between  prediction and experiment for the ratio would be a clear 
evidence for 
annihilation or other non-factorisable contributions. We find that  
annihilation topology  likely plays an 
indispensable role at least for penguin-dominated 
$\rm PV$ channels. Including the annihilation terms in QCDF, we find that
 the direct CP asymmetry of $B \to \pi^+ \pi^-$ to be 
about 3 times larger than that of $B \to K^\mp\pi^\pm$, with opposite sign,
in agreement with experiment\cite{Belle}. 

\section{QCD factorization for charmless $B$ decays}

 The effective Lagrangian for non-leptonic $B$ decays
can be obtained from  operator product expansion and 
renormalization group equation, in which short-distance effects
involving large virtual momenta of the loop corrections from the scale
$M_W$ down to $\mu ={\cal O}(m_b)$ are  integrated into the Wilson
coefficients. The amplitude for the decay $B \to M_1 M_2$ can be
expressed as
\be
\kern -0.1cm {\cal A}(B \rightarrow M_1 M_2)=
 \frac{G_F}{\sqrt{2}} \sum_{i=1}^6 \sum_{q=u,c}
\lambda_q C_i (\mu) \langle M_1 M_2 \vert O_i (\mu) \vert B \rangle
\label{BMM1}
\ee
 $\lambda_q$ is a CKM factor, $C_i (\mu)$ are the Wilson coefficients
perturbatively calculable from first principles and  $O_{i}$ are the
 tree and penguin operators given by(neglecting other operators):
\bea
&&O_{1}= (\bar{s}u)_{L}(\bar{u}b)_{L} \quad, \ O_{4}= \sum_{q}(\bar{s}q)_{L}
(\bar{q}b)_{L}\nonumber \\
&&O_{6}= -2\sum_{q}(\bar{s}_{L}q_{R})(\bar{q}_{R}b_{L})
\label{Oi}
\eea 
The   hadronic matrix elements
$\langle M_1 M_2 \vert O_i (\mu) \vert B \rangle$  contain
 the physics effects from the scale 
$\mu ={\cal O}(m_b)$ down to $\Lambda_{\rm QCD}$.
 In the heavy quark limit, QCD Factorisation~\cite{QCDF} allows 
the  decay mplitude
$\langle M_1 M_2 \vert O_i (\mu) \vert B \rangle$ to  be factorized 
into hard radiative corrections
and  non-perturbative matrix elements which can be parametrized by the
semi-leptonic decays form factors 
and meson light-cone distribution amplitudes (LCDAs). 

 Power corrections in $1/m_{b}$ come from penguin
matrix elements, chirally enhanced corrections and annihilation 
contributions. For example, in the $B \to \pi K$ amplitude, the 
matrix element of  $O_{6}$ is of the order $O(1/m_{b})$ compared to 
the $(V-A)\times (V-A)$ $O_{1}$ and $O_{4}$ matrix elements, since 
 $<K|\bar{s}_{L}d_{R}|0>$
is proportional to $m_{K}^{2}/m_{s}\approx 2.5\,\rm GeV$ while 
$<K|\bar{s}_{L}d_{L}|0>$ is proportional to  $K$ momentum which
is $O(m_{b})$, thus  numerically, the matrix element of $O_{6}$ 
has a factor
\be
r_\chi^K = \frac{2 m_K^2}{m_b (m_s + m_d)} \approx O(1)
\ee
and is comparable to that of $O_{4}$. For penguin-dominated decays, 
the $O_{4}$ and  $O_{6}$ matrix element are of the same sign in 
PP channnel, while in PV channel they are of opposite sign. 
Thus in QCDF one expects a small $B \to K\rho$ branching ratio 
relative to $B \to \pi K$. Because of a cancellation between the 
$O_{4}$ and $O_{6}$ contributions, the $B \to K\rho$ decay
is more sensitive to other power corrections and  non-factorisable 
contributions. Including the chirally-enhanced corrections
in terms of  two quantities $X_{\rm A,H}$ and 
a strong phase, the   $B \to M_1 M_2$ decay amplitudes in QCDF
can be thus be written as\cite{QCDF1,QCDF2}:
\bea
 \kern -0.2cm {\cal A}(B \rightarrow M_1 M_2)=
 \frac{G_F}{\sqrt{2}}\sum_{p=u,c}V_{pb}V^{*}_{ps}  
 \left( -\sum_{i=1}^6 a_i^p
   \langle M_1 M_2 \vert O_i \vert B \rangle_f +
 \sum_{j} f_B f_{M_1}f_{M_2} b_j \right ),
\label{BMM}
\eea

\section{Power corrections in $B \to \rm PV$ decays}

Consider the  ratio of $A(B^+\to \pi^+ K^{\ast 0})$ to 
$A(B^0 \to \rho^+ \pi^-)$. amplitudes. If the power
corrections were negligible, this ratio would be theoretically very
clean where the form factors cancel out, furthermore it is
almost independent of the  CKM angle $\gamma$ and the strange-quark mass:
 \bea \label{pik}
 \left \vert \frac{{\cal A}(B^+ \rightarrow \pi^+ K^{\ast 0})}{{\cal A}(B^0 \rightarrow
\rho^+ \pi^-)}\right \vert \simeq 
 \left \vert \frac{ V_{cb}V_{cs}}{V_{ub}V_{ud}} \right \vert
\frac{f_{K^\ast}}{f_\rho} \left \vert 
\frac{a_4^c (\pi K^\ast)+r_\chi^{K^\ast} a_6^c (\pi K^\ast) }{a_1^u}
\right \vert 
\label{rhopi}
\eea

 $\vert (a_4^c (\pi K^\ast)+r_\chi^{K^\ast} a_6^c (\pi
K^\ast) ) /a_1^u \vert$ should be about or less than 0.04 in QCDF.
($f_{K^\ast}/f_\rho \approx 1$). The ratio 
 $\vert V_{ub}/V_{cb}\vert $  is not very well determined
 experimentally, but a stringent lower limit can be obtained 
from the unitarity of the
CKM matrix . Since \cite{Pham,Buras} :
\begin{equation}\label{inequality}
\left \vert \frac{V_{ub}}{V_{cb}}\right \vert = \lambda \sin \beta 
\sqrt{1+\frac{\cos^2 \alpha}{\sin^2 \alpha}} \ge \lambda \sin \beta ~.
\end{equation} 
and from the current Babar and Belle measured values~:
$\sin 2\beta= 0.725 \pm 0.037 $ \cite{Smith}~, we have
\begin{equation}\label{inequality2}
 \left \vert \frac{V_{ub}}{V_{cb}} \right \vert \ge \lambda \sin \beta =
0.090 \pm 0.007 > 0.078 
\end{equation}
Eq.(\ref{rhopi}) implies the following inequality~:
\be\label{pikrhopi}
 0.53 > {\left \vert \frac{{\cal A}(B^+ \to \pi^+ K^{\ast
0})}{{\cal A}(B^0 \to \rho^+ \pi^-)}\right \vert }= 0.77 \pm 0.09~,
\ee
where the number on the rhs is from the measured branching ratios
\cite{HFAG,HFAG04}. The lhs
would be reduced further to $0.46 \pm 0.04$, if in 
Eq.(\ref{inequality}) one neglects a small $\cos^2 \alpha$ term according
to a  recent determined value $\alpha = (101^{+16}_{-9})^{\circ}$
\cite{Smith}.

\begin{figure}[htb]
\centering
\leavevmode
\epsfxsize=8cm
\epsffile{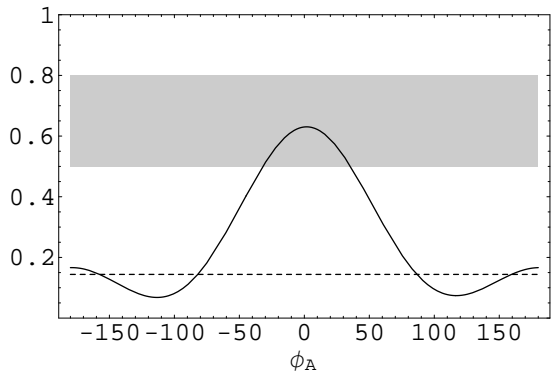}
\caption{ The ratio ${\cal B}(B^+ \to \pi^+ K^{\ast 0})/  
 {\cal B}(B^0 \to \rho^+ \pi^-)$ 
versus the weak annihilation phase
$\phi_A$. The default parameters are used but letting
the annihilation parameter $\rho_A=1$. The dashed lines show the
ratios without weak annihilation contributions. The gray areas denote
the experimental measurements with $1 \sigma$ error. }
\end{figure} 

\begin{figure}[htb]
\centering
\leavevmode
\epsfxsize=8cm
\epsffile{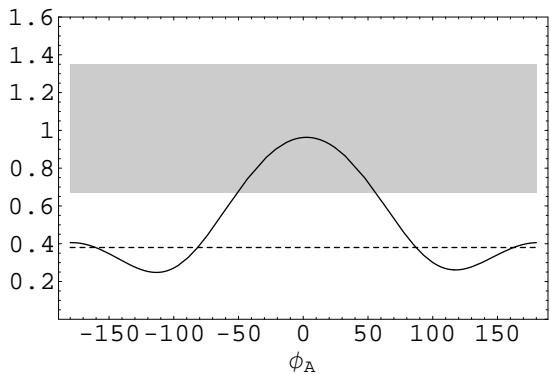}
\caption{ The ratio  
${\cal B}(B^0 \to K^+ \rho^-)/{\cal B}(B^0 \to \rho^- \pi^+)$
versus the weak annihilation phase
$\phi_A$. The default parameters are as in Fig.1 }
\end{figure} 

Since the  chirally enhanced corrections for penguin-dominated decays are
 not expected to be large, this large discrepancy 
 is   strong indication
that annihilation topology and/or other sources of power corrections might 
play an important role at least in $B \to \rm PV$ decays. There is  
 similar disagreement between theory and experiment in another ratio,
the branching fraction of $B^0 \to K^+ \rho^-$ to that of 
$B^0 \to \rho^- \pi^+$, though with large theoretical uncertainties.
For $\gamma=70^\circ$, $V_{ub}/V_{cb}=0.09$,
$a_4^c (\rho K)-r_\chi^K a_6^c(\rho K)=0.037+0.003 i$, $m_{s} = 90\,\rm MeV$,
we find 
\be
\frac{{\cal B}(B^0 \to K^+ \rho^-)}{{\cal B}(B^0 \to \rho^- \pi^+)} = 0.38
\ee
far below the measured value of  $1.01 \pm 0.34$, though, this 
ratio could be increased to $0.69$, if $m_{s}$ is lowered to 
$ 70\,\rm MeV$. The discrepancy could be greater according to a recent 
Belle measurements\cite{HFAG} which give a very large 
$B^0 \to K^+ \rho^-$ branching ratio,
$(15.1^{+3.4+2.4}_{-1.5-2.6})\times 10^{-6}$ which could  be obtained
by a large annihilation contributions\cite{QCDF2}. Recent charmed
meson inelastic rescattering FSI calculation\cite{Isola1} also produces 
a large branching ratio. Taken together, these results indicate that
the penguin-dominated  $B \to \rm PV$
decay amplitudes are consistently underestimated without annihilation 
contributions. Including the annihilation terms, from 
Eq. (\ref{BMM}), we have
\bea
   A(B^+ \rightarrow \pi^+ K^{\ast 0}) &=& f_{K^\ast}F^{B \pi}m_B^2 a_4 
+b_3(V,P)\nonumber \\
A(B^0 \rightarrow K^+ \rho^-) &=& f_K A^{B\rho}_0 m_B^2
(a_4-r_\chi^K a_6) + b_3(P,V)\kern 0.3cm
\eea
\be
  b_3 (M_1, M_2) = \frac{C_F}{N_c^2}\{ C_3 A_1^i (M_1,M_2)+   
 C_5 A^i_3 (M_1, M_2) +\kern -0.1cm (C_5+N_c C_6) A^f_3 (M_1, M_2) \}
\ee
With  the penguin
terms $a_4 \simeq -0.03$ and $a_4-r_\chi^K a_6 \simeq 0.037$ having
opposite sign, the key observation is that $b_3 (V, P)$ and $b_3(P, V)$ ,
which get most of the contribution from $(C_5+N_c C_6) A^f_3$ term, 
are also roughly of the opposite sign since  $A^f_3(P,V)=-A^f_3(V,P)$.
Thus QCDF can easily enhance both ratios without fine
tuning ( no large strong phase ) as can be seen in Fig.2 .

\section{Direct CP violations} 
We now turn to the CP asymmetries in QCDF with annihilation terms included.
Because of the CKM factor and $SU(3)$ symmetry for the tree and
penguin matrix elements in   $B^{0} \to\pi^+ \pi^- $ and 
$B^{0}\to K^{+}\pi^-$ decays, one can derive a relation between 
direct CP asymmetries in these two channels. With the CP asymmetry given as:
\begin{eqnarray}
&&A_{\pi \pi}  =   
 \frac{4 \vert V_{ub}V_{ud}
V_{cb}V_{cd} T_{\pi\pi} P_{\pi\pi}\vert \sin \gamma \sin \delta }
{2 {\cal B}(B \rightarrow\pi^+ \pi^- )} ~\mbox{,} \nonumber \\
&&A_{\pi K} = -\frac{4 \vert V_{ub}V_{us} 
V_{cb}V_{cs} { T}_{\pi K} { P}_{\pi K}\vert \sin \gamma \sin {\tilde \delta} }
{2 {\cal B}(B \rightarrow\pi^+ K^- )}. 
\label{ACP}
\end{eqnarray} 
($\delta=\delta_P - \delta_T$ = strong phases difference between
the penguin and tree amplitudes), we find
\bea\label{DCPV}
 \frac{A_{\pi\pi}}{A_{\pi K}}&=&-\frac{f_\pi^2}{f_K^2}
\frac{{\cal B}(B \to\pi^+ K^- )}
{{\cal B}(B \to\pi^+ \pi^- )} \left \vert \frac{T_{\pi\pi}P_{\pi\pi}}
{{ T}_{\pi K} {P}_{\pi K}} \right \vert 
\frac{\sin \delta}{\sin {\tilde \delta}} \nonumber \\
& \simeq & (-2.7 \pm 0.3) 
\frac{\sin \delta}{\sin {\tilde \delta}} 
\eea
a consequence of the fact that  
$T_{\pi\pi}P_{\pi\pi}/{ T}_{\pi K} {P}_{\pi K}$ is close to $1$,
a reasonable approximation in QCDF, at about $10$ percent level uncertainty.
A previous derivation of this relation is given in \cite{Fleischer,Dariescu}.
Belle has claimed large direct CP asymmetry
observed in $B^0 \to \pi^+ \pi^-$ decay while BaBar has not confirmed
it yet, but both of them are close in measurements on $A_{CP}(\pi^- K^+)$
\cite{Belle,BaBar,HFAG05}
\be
A_{\pi\pi}=\left \{ \begin{array}{ll}
0.56 \pm 0.12 \pm 0.06 & \mbox{(Belle)}~, \\
0.09 \pm 0.15 \pm 0.04 & \mbox{(BaBar)}~,\\
0.37 \pm 0.10   & \mbox{(Average)}~. 
\end{array} \right. 
\ee

\be
A_{\pi K}=\left \{ \begin{array}{ll}
-0.101 \pm 0.025 \pm 0.005 & \mbox{(Belle)}~, \\
-0.133 \pm 0.030 \pm 0.009 & \mbox{(BaBar)}~, \\
-0.114 \pm 0.020   & \mbox{(Average)}.~ 
\end{array} \right. 
\ee
We thus  expect  very 
naturally a larger direct CP violation
for $\pi^+ \pi^-$ decay compared with $\pi^- K^+$ decay, since 
the $\pi^+ \pi^-$ decay rate is smaller than
the $\pi^- K^+$ decay rate by  factor $3-4$, 

Experimentally, 
\begin{equation}
\frac{A_{\pi\pi}}{A_{\pi K}}=\frac{0.37 \pm 0.10}{-0.11 \pm 0.03}
  =-3.4 \pm 1.5 ~,
\end{equation}
still consistent with the theoretical estimation of $-2.7 \pm 0.3$.

Similar relation between CP asymmetries for the $B \to PV$ decays
for which the CP-violating interference terms are essentially of the same
magnitude, but with opposite sign:
\begin{eqnarray}
&&\kern -0.8cm\frac{A_{\rm CP}(B^0\kern -0.1cm \to \kern -0.1cm\rho^+ \pi^-)}{A_{\rm CP}(B^0\kern -0.1cm \to \kern -0.1cm  K^{\ast +} \pi^-)}
\simeq  -\frac{{\cal B}(B^0\kern -0.1cm\to \kern -0.1cm  K^{\ast +} \pi^-)}{{\cal B}(B^0\kern -0.1cm \to \kern -0.1cm
\rho^+ \pi^-)}\frac{f_\rho^2}{f_{K^\ast}^2}\frac{\sin \delta_{\pi \rho}}
{\sin \delta_{\pi K^\ast}} \nonumber \\
&&\kern -0.8cm\frac{A_{\rm CP}(B^0\kern -0.1cm \to \kern -0.1cm \rho^- \pi^+)}{A_{\rm CP}(B^0\kern -0.1cm \to \kern -0.1cm \rho^- K^+ )}
\simeq  -\frac{{\cal B}(B^0 \kern -0.1cm\to \kern -0.1cm
  \rho^- K^+ )}{{\cal B}(B^0 \kern -0.2cm\to \kern -0.2cm 
\rho^- \pi^+)}\frac{f_\pi^2}{f_{K}^2}\frac{\sin \delta_{\rho \pi}}
{\sin \delta_{\rho K}}
\end{eqnarray}
In the presence of charming penguin or charmed meson 
inelastic rescattering FSI effects, the above CP asymmetry relation
applies since the tree-penguin interference terms are related by $SU(3)$ 
symmetry and CKM factor\cite{Isola}. Thus any significant deviation 
from theoretical estimation would suggest either different 
strong phases, e.g  between $\pi \pi$ and $\pi K$ decays,  or possible 
new physics contributions.

\section{Conclusion}

 Power corrections in charmless B
decays are probably large, at least for the penguin-dominated PV channel.
 The key observation is that QCDF predicts 
the annihilation terms for $B^+\to \pi^+ K^{\ast 0}$ and $B^0 \to K^+ \rho^-$
to be   almost equal in  magnitude but opposite in sign and thus
enhance the decay rates for these two modes  to accommodate 
the experimental data. The relation for the direct CP asymmetry would 
naturally implies a large CP asymmetry
for $B \to \pi^+ \pi^-$ ,  about 3 times larger than that of
$B \to \pi^\pm K^\mp$ with opposite sign.

\bigskip

I would like to thank G. Nardulli, P. Colangelo, F. De Fazio  and 
the organisers of QCD@Work for the
warm hospitality extended to me at Conversano.


\begin{thebibliography}{99}

\bibitem{BaBar}
BaBar Collaboration, B.Aubert {\it et al}, hep-ex/0501071.

\bibitem{Belle}
Belle Collaboration, K. Abe {\it et al.}, hep-ex/0502035.

\bibitem{Isola} C. Isola, M. Ladisa, G. Nardulli, T. N. Pham  and P. 
Santorelli, Phys. Rev. D {\bf 64}, 014029 (2001); ibid, {\bf 65}, 
094005 (2002).

\bibitem{Isola1} C. Isola, M. Ladisa, G. Nardulli, and P. Santorelli, 
Phys. Rev. D {\bf 68}, 114001 (2003) .

\bibitem{Ciuchini} M. Ciuchini, E. Franco, G. Martinelli, and
L. Silvestrini, Phys. Lett. {\bf B 515}, 33 (2001) ; and previous works
cited therein.

\bibitem{QCDF}
M. Beneke, G. Buchalla, M. Neubert and C. T. Sachrajda, Phys. Rev. Lett.
{\bf 83}, 1914 (1999); Nucl. Phys. B {\bf 591}, 313  (2000) .

\bibitem{pQCD}
Y. Y. Keum, H. N. Li and A. I. Sanda, Phys. Rev. D {\bf 63}, 054008 (2001) ;
Phys. Lett. B {\bf 504}, 6 (2001) .

\bibitem{Du}
D. S. Du, J. F. Sun, D.S. Yang and G. H. Zhu, Phys. Rev. D {\bf 67}, 014023
(2003) . 

\bibitem{Aleksan}
R. Aleksan, P. F. Giraud, V. Morenas, O. Pene and A. S. Safir, Phys. Rev. D 
{\bf 67}, 094019 (2003) .

\bibitem{Cottingham}
N. de Groot, W. N. Cottingham and I. B. Whittingham, Phys. Rev. D {\bf 68}, 
113005 (2003).

\bibitem{Zhu} T. N. Pham and Guohuai Zhu, Phys. Rev. D {\bf 69} ,
114016  (2004).

\bibitem{QCDF1}
M. Beneke, G. Buchalla, M. Neubert and C. T. Sachrajda, Nucl. Phys. B 
{\bf 606}, 245 (2001).

\bibitem{QCDF2}
M. Beneke and M. Neubert, Nucl. Phys. B {\bf 675}, 333 (2003) .

\bibitem{Pham}
T. N. Pham, invited talk at the 2nd Workshop on the CKM Unitarity
Triangle, Durham, April 2003, hep-ph/0306271 .

\bibitem{Buras}
A. J. Buras, F. Parodi and A. Stocchi, JHEP {\bf 0301}, 029 (2003) .

\bibitem{Smith}
J. Smith, Talk given at the 3rd Workshop on the CKM Unitarity
Triangle,  San Diego, USA, March 2005 .

\bibitem{HFAG} Heavy Flavor Averaging Group(HFAG), hep-ph/0505100 .

\bibitem{HFAG04}Heavy Flavor Averaging Group(HFAG),\\
{\tt www.slac.stanford.edu/xorg/hfag/triangle/ichep2004/index.shtml} .

\bibitem{HFAG05}Heavy Flavor Averaging Group(HFAG),\\
{\tt www.slac.stanford.edu/xorg/hfag/triangle/summer2005/index.shtml} .

\bibitem{Fleischer}  R. Fleischer, Phys. Lett. B {\bf 459} (1999) 306~;
A. J. Buras, R. Fleischer, S. Recksiegel and F. Schwab, hep-ph/0402112 . 

\bibitem{Dariescu}  M. Dariescu, N. G. Deshpande, X. G. He, and
G. Valencia,  Phys. Lett. B {\bf 557}, 60 (2003).

\end{thebibliography}
\end{document}